\documentclass[9pt,journal]{IEEEtran}
\pdfoutput=1
\usepackage{booktabs} 
\usepackage[pdftex]{graphicx}
\usepackage{amsmath}
\usepackage{enumitem}
\usepackage{setspace}
\usepackage{array}
\usepackage{etoolbox}
\usepackage{algorithmic}
\usepackage{subfig}
\usepackage[T1]{fontenc}
\usepackage{dblfloatfix}

\usepackage{cite}

\usepackage{url}

\makeatletter
\IEEEtriggercmd{\reset@font\normalfont\fontsize{6.8pt}{8pt}\selectfont}
\makeatother
\IEEEtriggeratref{1}

\makeatletter
\let\old@ps@headings\ps@headings
\let\old@ps@IEEEtitlepagestyle\ps@IEEEtitlepagestyle
\def\confheader#1{%
	\def\ps@headings{%
		\old@ps@headings%
		\def\@oddhead{\strut\hfill#1\hfill\strut}%
		\def\@evenhead{\strut\hfill#1\hfill\strut}%
	}%
	\def\ps@IEEEtitlepagestyle{%
		\old@ps@IEEEtitlepagestyle%
		\def\@oddhead{\strut\hfill#1\hfill\strut}%
		\def\@evenhead{\strut\hfill#1\hfill\strut}%
	}%
	\ps@headings%
}
\makeatother

\begin{document}

	\title{Fault-Tolerant Nanosatellite Computing on a Budget}
	
	\author{Christian~M.~Fuchs,~\IEEEmembership{Member,~IEEE,}
		    Nadia~M.~Murillo, Aske~Plaat, Erik~van~der~Kouwe, Daniel~Harsono,
		    and~Todor~P.~Stefanov,~\IEEEmembership{Member,~IEEE}\vspace{-20pt}
	\thanks{C.M. Fuchs was with the Leiden Institute of Advanced Computer Science and Leiden Observatory at Leiden University, 2333 CA, The Netherlands, e-mail: christian.fuchs@dependable.space}
	\thanks{A. Plaat, E.v.d. Kouwe, and T.P. Stefanov were with the Leiden Institute of Advanced Computer Science}
	\thanks{N.M. Murillo and D. Harsono were with Leiden Observatory}
	\thanks{This approach was developed for a 4-year European Space Agency (ESA) NPI project supported by two industrial partners.
		N.M. Murillo and D. Harsono acknowledge funding through the European Union A-ERC grant 291141 CHEMPLAN, by the Netherlands Research School for Astronomy (NOVA), and the Royal Netherlands Academy of Arts and Sciences Professor Prize.}
	\thanks{Manuscript submitted at RADECS2018 on April 17th, 2018, and revised, reworked and extended on September 16th.}}
	
	\markboth{}
	{Fuchs \MakeLowercase{\textit{et al.}}: Fault-Tolerant Nanosatellite Computing on a Budget}

	\maketitle
	
	\begin{abstract}
		We present an on-board computer architecture designed for small satellites ($<$50kg), which exploits software-fault-tolerance to achieve strong fault coverage with commodity hardware.
		Micro- and nanosatellites have become popular platforms for a variety of commercial and scientific applications, but today are considered suitable mainly for short and low-priority space missions due to their low reliability.
		In part, this can be attributed to their reliance upon cheap, low-feature size, COTS components originally designed for embedded and mobile-market applications, for which traditional hardware-voting concepts are ineffective.
		Software-fault-tolerance has been shown to be effective for such systems, but have largely been ignored by the space industry due to low maturity, as most have only been researched in theory.
		In practice, designers of payload instruments and miniaturized satellites are usually forced to sacrifice reliability in favor of delivering the level of performance necessary for cutting-edge science and innovative commercial applications.
		Thus, we developed a set of software measures facilitating fault tolerance based upon thread-level coarse-grain lockstep, which we validated through fault-injection.
		To offer strong long-term fault coverage, our architecture is implemented as tiled MPSoC on an FPGA, utilizing partial reconfiguration, as well as mixed criticality.
		This architecture can satisfy the high performance requirements of current and future scientific and commercial space missions at very low cost, while offering the strong fault-coverage guarantees necessary for platform control even for missions with a long duration.
		This architecture was developed for a 4-year ESA project. Together with two industrial partners, we are developing a prototype to then undergo radiation testing.
	\end{abstract}
	
	
	\section{Introduction}
	Satellite miniaturization has enabled a broad variety of scientific and commercial space missions, which previously were technically infeasible, impractical or simply uneconomical.
	However, due to their low reliability, nanosatellites, as well as light microsatellites, are typically not considered suitable for critical and complex multi-phased missions and high-priority science.
	The on-board computer (OBC) and related electronics constitute a large part of such spacecraft, and were shown to be responsible for a significant share of post-deployment failure \cite{langer2016reliability}.
	Indeed, these components often lack even basic fault tolerance (FT) capabilities.
	
	Due to budget, energy, mass, and volume restrictions, existing FT solutions originally developed for larger spacecraft can not be adopted.
	In this paper we describe an multiprocessor System-on-Chip (MPSoC) that utilizes conventional hardware, providing FT for miniaturized satellites.
	The MPSoC is assembled from well tested COTS components, library logic (IP), and powerful embedded and mobile-market processor cores, yielding a non-proprietary, open architecture.
	Our key contribution is a fault tolerant OBC architecture for CubeSat use that consists only of extensively validated standard parts, and can be reproduced with minimal manpower and financial resources.
	
	\section{Background \& Related Work}
	
	Aboard nanosatellites, subsystems are controlled by just one command \& data handling system, whereas aboard a larger satellite these tasks are distributed across multiple dedicated payload and subsystem computers.
	This implies a varying OBC workload throughout a nanosatellites mission, which traditional FT solutions only handle through over-provisioning.
	The tiled MPSoC design presented in this paper can efficiently handle faults through thread migration and partial reconfiguration.
	Major parts of our approach are implemented in software, allowing the OBC to deliver the desired combination of performance, robustness, functionality, or to meet a specific power budget.
	To enable strong FT with low-cost commodity hardware, we combine fault detection, isolation and recovery in software, FPGA configuration scrubbing with other fault detection, isolation and recovery (FDIR) measures across the embedded stack.

	Nanosatellites today utilize almost exclusively COTS microcontrollers and application processors-SoCs, FPGAs, and combinations thereof \cite{kastensmidt2016fpgas, carlson2016use}.
	Due to manufacturing in fine technology nodes, and the use of extensively optimized standard IP, they offer superior efficiency and performance as compared to space-grade OBC designs. 
	The energy threshold above which highly charged particles can induce faults (SEE -- single event effects) in such components decreases, while the ratio of events inducing multi-bit upsets (MBU), and the likelihood of permanent faults, increase.
	To adapt such hardware-FT based concepts additional FT-circuitry is required, inflating logic size and producing diminishing returns, resulting in limited scalability and low clock frequencies \cite{gupta2015shakti, pigno2011testbench, jackson2016implementation}.
	We can observe that traditional FT-concepts applied to modern COTS hardware yield no nanosatellite compatible architectures.

	While more sensitive to transient faults than ASICs \cite{berg2015nasaFPGAtestOverview, Tambara2015zynqGoodRadTestPerformance}, FPGA-based Soft-SoCs have been shown to offer excellent FDIR potential for miniaturized satellites \cite{Wirthlin2015HighReliabilityFPGA}.
	Transients in critical parts of the FPGA fabric can be scrubbed \cite{stoddard2017hybrid}, while permanent faults may be compensated through reconfiguration with differently routed configuration variants \cite{bozzoli2017self}.
	Fine-grained, non-invasive fault detection in FPGA fabric, however, is challenging, and subject of ongoing research \cite{ebrahimi2016low, rittner2017automated}.
	Relevant FT-concepts thus rely on error scrubbing, which has scalability limitations and cover only parts of the fabric \cite{ebrahimi2016low, stoddard2017hybrid}.
	We overcome these limitations by implementing fault-detection in software through thread-replication and coarse-grain lockstep within an MPSoC using weakly coupled cores.
	
	Tiled architectures \cite{singh2013mapping, meloni2012system} are often used for well paralellizable applications with many low-performance processor cores.
	Among others, \cite{beechu2017hardware} and \cite{meloni2012system} showed that such typologies can also be exploited to achieve FT for image processing applications with a very specific structure.
	We combine a tiled architecture with coarse-grained lockstep \cite{fuchs2017atATS}, enabling FDIR without constraining the application type or system architecture.
	Thus, the architecture presented in this paper is well suited for platform control and can be used as a template, allowing a high level of OBC design freedom, and enabling a considerable amount of testing to be inherited from COTS components and logic.
	
	Thread migration has been shown to be a powerful tool for assuring FT, but prior research ignores fault detection, and imposed tight constraints on an application's type and structure (e.g., video streaming and image processing \cite{martinez2017fully}).
	Thread-level coarse-grain lockstep of weakly coupled cores instead supports general purpose computing, and in the past, has already been used for high availability, non-stop service, and error resilience concepts.
	However, in prior research, faults are usually assumed to be isolated, side effect free, and local to an individual application thread \cite{holler2015software} or transient \cite{dobel2014operating, munk2015toward}, entailing high performance \cite{santangelo2013open} or resource overhead \cite{missimer2014distributed, al2016fault}.
	More advanced proof-of-concepts \cite{kretzschmar2016synchronization, dobel2014operating}, however, attempt to address these limitations, and even show a modest performance overhead between 3\% and 25\%, but utilize checkpoint \& rollback or restart mechanics \cite{dobel2014operating}, which make them unsuitable for spacecraft command \& control applications.
	
	Many of these limitations and obstacles ultimately can be attributed to low maturity, as a majority of software-FT concepts are published as a concept TRL1 but remain unvalidated.
	Hence, they could be uncovered, and in many cases, can be potentially resolved through implementation and practical validation \cite{kretzschmar2016synchronization}, increasing maturity to TRL2 or TRL3.
	However, development of a testable proof-of-concept is a time consuming and costly undertaking \cite{natella2016assessing}, as outlined among others by Sangchoolie et al. \cite{sangchoolie2017light} with limited immediate yield for academic publication.
	Fault injection for entire OS instances is especially non-trivial \cite{cotroneo2012experimental}, as thorough preparation and careful tool-selection is necessary to obtain representative results from a fault injection experiment \cite{natella2013fault}.
	Therefore, a broad variety of TRL1 software-FT concepts exist today at a theoretical level \cite{malik2011adaptive, smiri2016fault, al2016four}, for which validation was only conducted statistically using modeling with different fault distributions or not a all.
	In this contribution, we therefore conduct validation of our coarse-grain lockstep approach using systematic fault-injection.
	Thereby we verify the effectiveness of our coarse-grain lockstep FDIR mechanics under stress using a RTOS-based proof-of-concept implementation, increasing maturity to TRL3.
	
	\section{A Hybrid Fault-Tolerance Approach}
	\label{sec:stageoverview}
\begin{figure}[!b]
		\vspace{-10pt}
		\centering
		\includegraphics[width=0.9\linewidth]{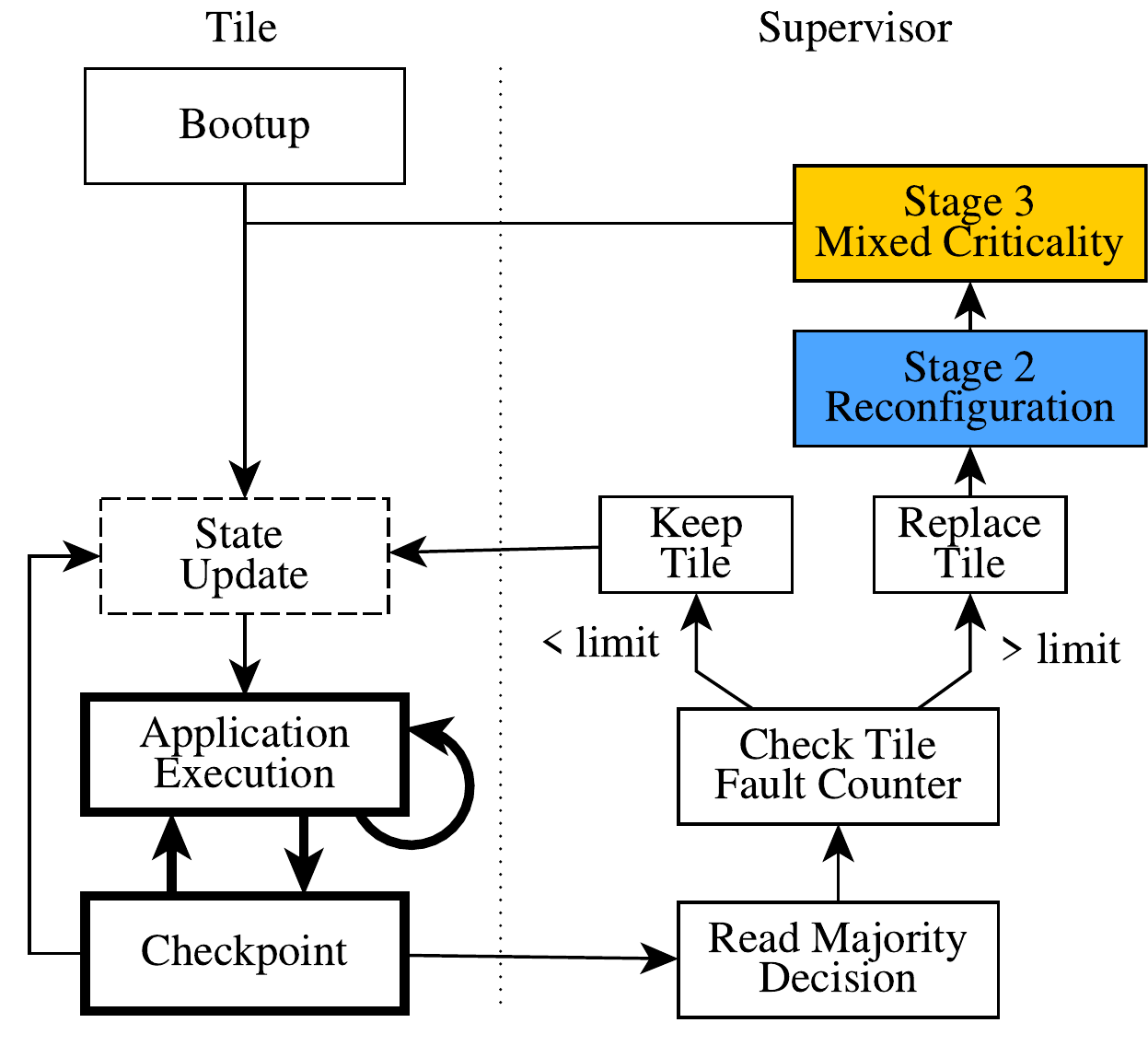}
		\caption{Stage 1 (white) assures fault detection (bold) and fault coverage. Stages 2 (blue) and 3 (yellow) counter resource exhaustion and adapt the on-board computer application schedule to reduced system resources.}
		\label{fig:supervision}
	\end{figure}

	Conventional FT architectures require proprietary logic in hardware to facilitate fault detection and coverage.
	In contrast, the architecture described in this paper can offer strong FT using just COTS components and proven standard library logic.
	This is made possible through the use of the FT approach we presented in \cite{fuchs2017atATS}.
	The high-level functionality of this approach is depicted in Fig. \ref{fig:supervision}, and consists of three interlinked fault mitigation stages implemented across the embedded stack:

	\textbf{Stage 1} implements forward error correction and utilizes coarse-grain lockstep of weakly coupled cores to generate a distributed majority decision across tiles.
	Fault detection is facilitated through application callback functions, without requiring deep modifications to an application or knowledge about intrinsics.
	
	\textbf{Stage 2} recovers failed tiles through reconfiguration and self-testing.
	It assures the integrity of programmed logic and deploys configuration scrubbing, as well as Xilinx Soft-Error-Mitigation (SEM), to correct transients in FPGA fabric.
	Its objective is to assure and recover the integrity of processor cores and their immediate peripheral IP through FPGA reconfiguration and the use of differently routed and placed alternative configuration variants, thereby counteracting resource exhaustion.
	
	\textbf{Stage 3} engages when too few healthy tiles are available, and re-allocates processing time to maintain reliability.
	To do so, thread-level mixed criticality is exploited, assuring sufficient compute resources are available to high-criticality applications by sacrificing performance or availability of lower-criticality threads.
	
	Further details including benchmark results are available in \cite{fuchs2017atATS}.
	The main target in our project is the ARM Cortex-A53 application processor, which is today widely used in embedded and mobile-market devices.
	However, this research is processor and ISA independent.
	In this paper, we describe an MPSoC design and architecture template, which is enabled by this approach and can be reproduced in Xilinx Vivado 2017.1 and later.

	\subsection*{Stage 1: Short-Term Fault Mitigation}
	\label{sec:stage1}
	The objective of Stage 1 is to detect and correct faults within a tile, and assure a consistent system state through checkpoint-based FEC.
	It is implemented as sets of tiles running two or more copies of application threads (siblings) in lock step.
	Checkpoints interrupt execution, facilitating the lockstep and enforcing synchronization, allowing thread assignment within the system to be adjusted if required.

	This approach enables us to utilize application intrinsic code and data to assess the health state of the system without requiring in-depth knowledge about the application itself.
	The supervisor reads out the results of the tiles' decentralized consistency decision.
	Application threads can be scheduled and executed in an arbitrary order between two checkpoints, as long as their state is equivalent upon the next checkpoint.
	
	We avoid thread synchronization issues as encountered by Kretzschmar et al. in \cite{kretzschmar2016synchronization} by merely reusing existing OS functionality without breaking or ABI/API guarantees.
	Therefore, we can continue relying upon pre-existing synchronization mechanics such as POSIX cancellation points\footnote{For example, sleep, yield, pause; for further details, see IEEE Std 1003.1-2017 p517} and their bare-metal equivalents (e.g., \emph{RTEMS\_NO\_PREEMPT} in RTEMS's Classic API if used instead of \emph{newlib} or the POSIX API).
	
	Stage 1 can deliver real-time guarantees if required, and the tightness of the RT guarantees depends upon the time required to execute application callbacks.
	In our RTEMS/POSIX-based implementation, we utilize priority-based, preemptive scheduling with timeslicing, allowing threads to delay checkpoints until they reach a viable state for checksum comparison.
	
	An application should provide four callback routines to the OS, which are executed during tile boot by the OS or as part of a checkpoint routine:
	\begin{itemize}[leftmargin=.35cm]
		\item an \emph{initialization routine}, to be executed on all tiles at bootup;
		\item a \emph{checksum callback}, used to generate a checksum for comparison with siblings,
		\item a \emph{expose state callback}, exposing all thread-state relevant data to synchronize a sibling with a lockstep group;
		This data can either be placed directly in the tile's local memory, or as a reference to structures in main memory.
		\item an \emph{update callback}, which is executed on a tile that needs to synchronize its state to a lockstep group.
	\end{itemize}

	Besides the addition of these callbacks, no alterations to an application's logic are necessary, except a viable way to assure it can be interrupted by a callback routine periodically.
	The required development effort for implementing these features in general is comparably low, but depends on the structure of an application.
	For the astronomical instrumentation applications utilized in our proof-of-concept, these routines could be implemented with 10-20 lines of C-code each.
	For example, the checksum callback consists almost exclusively of CRC library calls for generating a checksum from a set of state relevant variables and data structures in heap and stack.
	
	\begin{figure}[!b]
		\vspace{-15pt}
		\centering
		\includegraphics[width=0.75\linewidth,clip, trim = 0pt 0pt 0pt 0pt]{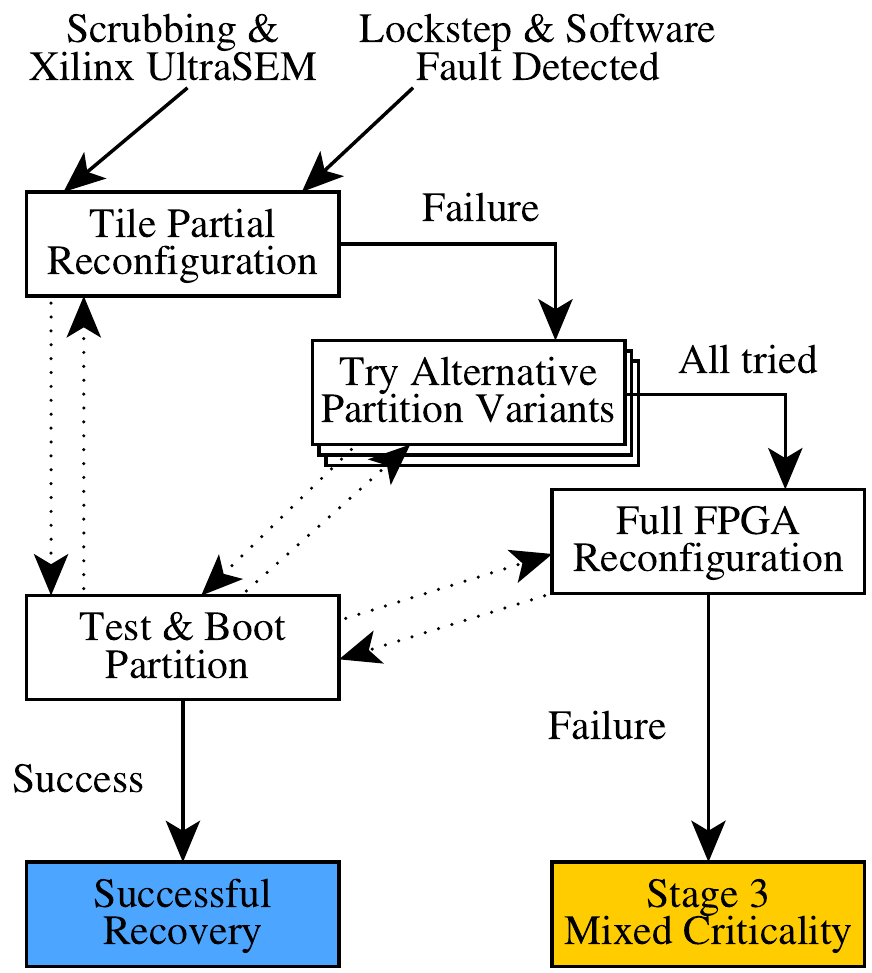}
		\caption{The objective of Stage 2 is to recover defective tiles and other logic through partial and full FPGA reconfiguration via ICAP.
			If this is unsuccessful as well and no further spare processing capacity is available to handle future faults, Stage 3 is activated to find a more resource conserving application schedule, replenishing the spare resource pool.
		}
		\label{fig:stage2}
	\end{figure}

	Callbacks may be omitted due to practical reasons.
	For applications which require little code and time for initialization, the initialization routine can be omitted.
	Applications which are not executed continuously could return a pre-generated checksum to the OS, instead of providing checksum, synchronization and callback handlers, for example, by providing the OS with a signature or checksum before program termination.
	Applications without a persistent state, or in which the state is continuously re-generated based on input data, no update callback would be necessary.
	
	Checkpoints were designed to be time triggered on each tile independently, but can also be induced by the supervisor through an interrupt, for example, to signal that new threads have been assigned (see also Section \ref{sec:applications} for additional information on time-vs-interrupt driven checkpoint triggering).
	Thus, the OS only has to support interrupts, timers, and a multi-threading-capable scheduler.
	To the best of our knowledge, such functionality is available in all widely used RT- and general purpose OS implementations.
	
	\subsection*{Stage 2: Tile Repair \& Recovery}
	\label{sec:stage2}
	Stage 1 can not reclaim defective tiles, eventually resulting in resource exhaustion.
	Therefore, in this stage, we recover defective tiles through reconfiguration to counter transients in FPGA fabric.
	To do so, the supervisor will first attempt to recover a tile using partial reconfiguration.
	Afterwards, the supervisor validates the relevant partitions to detect permanent damage to the FPGA (well described in, e.g., \cite{nguyen2017repairing}), and executes self-test functionality on the tile to detect faults in the tile's main memory segment and peripherals.
	If unsuccessful, the supervisor will repeat this procedure with differently routed configuration variants, potentially avoiding or repurposing permanently defective logic.
	
	The supervisor can also attempt full reconfiguration implying a full reboot of all tiles.
	Further details on reconfiguration and error scrubbing with a microcontroller-based proof-of-concept implementation for a nanosatellite are available in \cite{fuchs2016enhancing}.
	If both partial- and full-reconfiguration are unsuccessful and all spare resources have been exhausted, Stage 3 is utilized to assure a stable system core to enable operator intervention.
	
	\begin{figure}[!b]
		\centering
		\vspace{-20pt}
		\includegraphics[width=0.9\linewidth,clip, trim = 0pt 0pt 0pt 0pt]{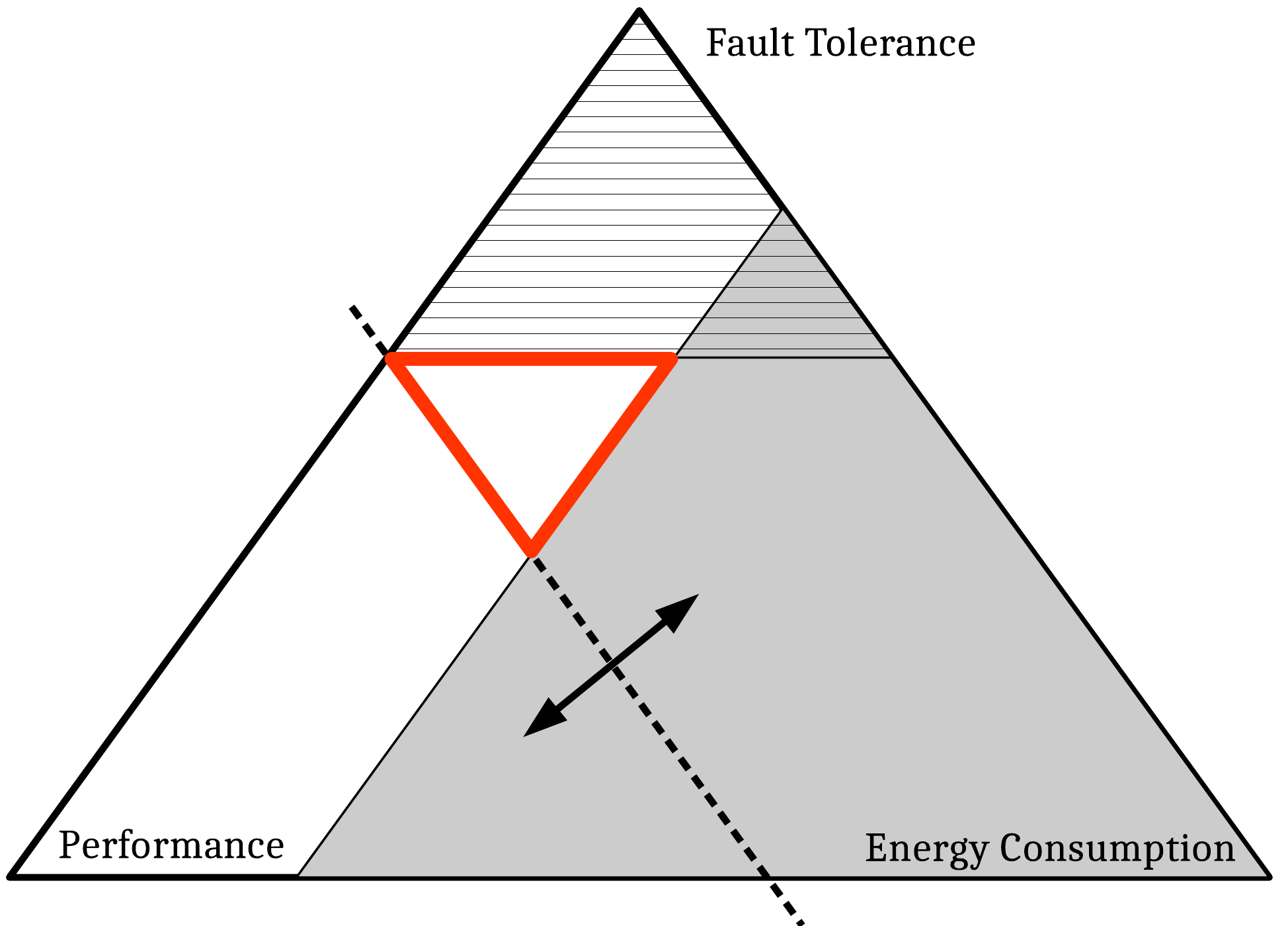}
		\caption{Our architecture allows the system properties of fault-tolerance, performance, and energy consumption of an OBC to be adjusted at runtime. 
		The spacecraft operator can prioritize one of these objectives, e.g. to achieve minimum energy consumption over computational performance, while maintaining a given level of fault tolerance.
		}
		\label{fig:FtSpeedEnergyTriangle}
	\end{figure}

	\begin{figure*}[!b]
	\centering
	\includegraphics[width=0.85\linewidth,clip, trim = 0pt 0pt 0pt 0pt]{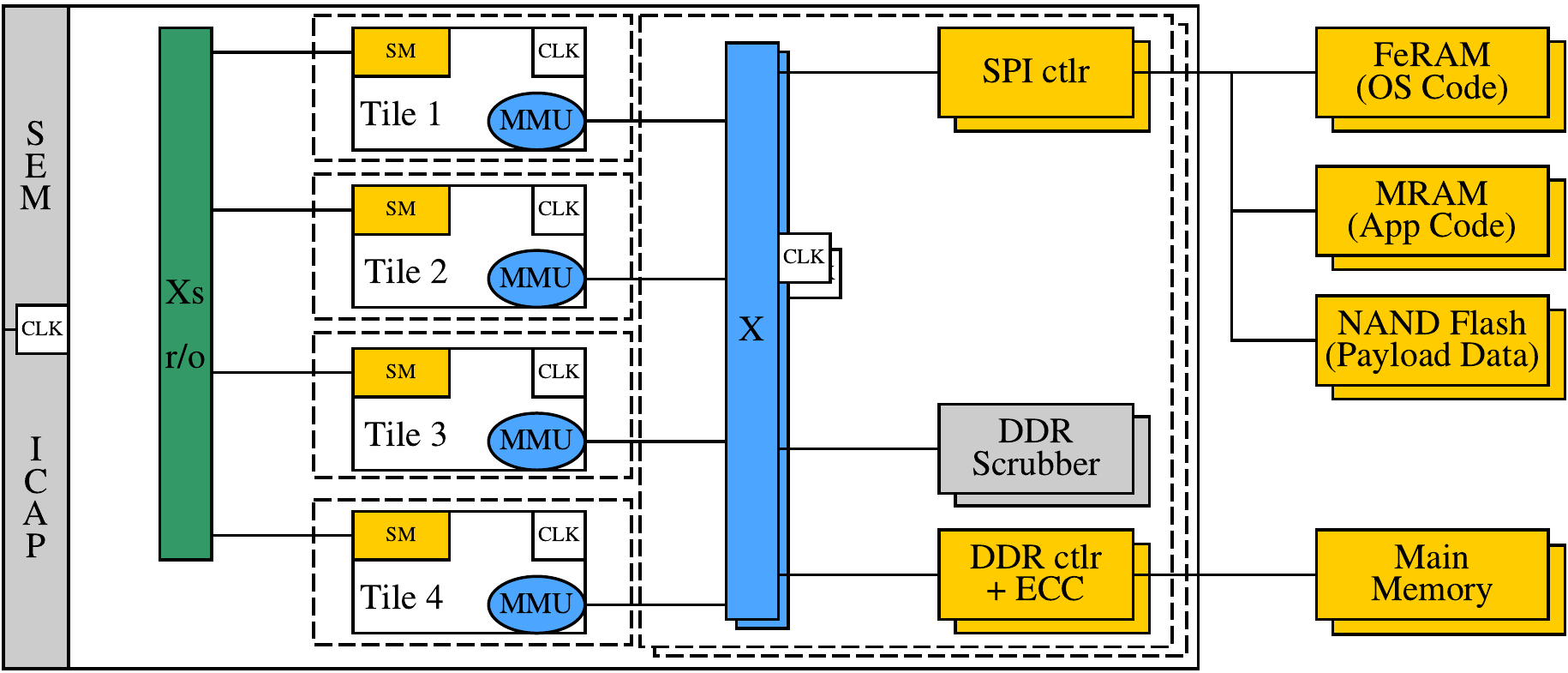}
	\caption{The topology of our tiled MPSoC design.
		Each tile exists in its own reconfiguration partition and therefore also clock domain, simplifying routing and logic placement.
		Reconfiguration partitions are indicated with dashed lines.
	}
	\label{fig:mpsoc}
	\end{figure*}
	\subsection*{Stage 3: Applied Mixed Criticality}
	\label{sec:stage3}
	Stage 3 maintains system stability of an aged or degraded OBC, if the remaining healthy tiles of the MPSoC no longer have sufficient processing capacity available for all applications.
	When considering a miniaturized satellite's OBC, we can differentiate individual applications and parts of the flight software by criticality.
	At the very least, we will find software essential to a satellite's operation, for example, platform control and commandeering, as well as other applications of various levels of lower criticality.
	If the previous stages no longer have enough spare processing capacity or tiles to compensate a fault, this stage utilizes thread-level mixed criticality to assure stability of core OBC functions.
	To do so, it can sacrifice lower criticality tasks in favor of providing compute resources to reach the desired replication level for critical threads.
	
	Dependability for higher-criticality threads can be maintained efficiently by reducing compute performance or reliability of lower-criticality applications.
	Lower-criticality tasks may be executed less frequently or on fewer tiles, thereby reducing functionality or fault coverage for these tasks, retaining resources for higher-criticality threads.
	This decision is taken autonomously, and the operator can then define a more resource conserving satellite operation schedule at the spacecraft level (e.g., sacrifice link capacity, or on-board storage space) to make the best use of the OBC in its degraded state.
	
	In practice a satellite operator can use this functionality also to dynamically adjust the performance of the MPSoC mid mission.
	This is achieved by adapting the distribution of applications across tiles, the level of replication of application threads, and the processing time allocated to individual application threads.
	The three properties, thus, are in competition to eachother, as depicted in Figure \ref{fig:FtSpeedEnergyTriangle}.
	This capability is analogous to the powersaving capabilities present in today's mobile devices and consumer desktop computers, where performance and energy consumption objective compete.
	An optimal combination of these objectives exists only in theory, but in practice would be very costly to obtain.
	For practical use, a set of ``good enough but non-optimal'' can be achieved as at runtime autonomously using heuristics.
	Further information on Stage 3 including dynamic thread-mapping, as well as performance, energy and robustness optimization at run-time is available in \cite{fuchs2018atAHS}.

	\section{The MPSoC Architecture}
	\label{sec:mpsoc} 
	We developed our software-FT architecture for use on top of an MPSoC consisting only of COTS technology.
	The main target in our project is the ARM Cortex-A53 application processor.
	For many size-optimized space applications, smaller cores such as the Cortex-A32, A35 and A5 may also offer a better balance between performance, universal platform support, and logic utilization.
	The Cortex-A53 core was chosen as it is today widely used in a variety of industrial and mobile-market devices, though our architecture is processor and instruction set architecture (ISA) independent.
	
	In this section, we describe a publicly reproducible MPSoC design variant implementing our architecture, which can be designed in full using Xilinx library IP and Microblaze processor cores.
	The architecture minimizes shared logic, compartmentalizes tiles, and offers a clearly defined access channel between tiles and the supervisor, and is depicted in Figure \ref{fig:mpsoc}.

	\begin{figure*}
		\centering
		\includegraphics[width=0.9\linewidth,clip, trim = 10pt 0pt 0pt 0pt]{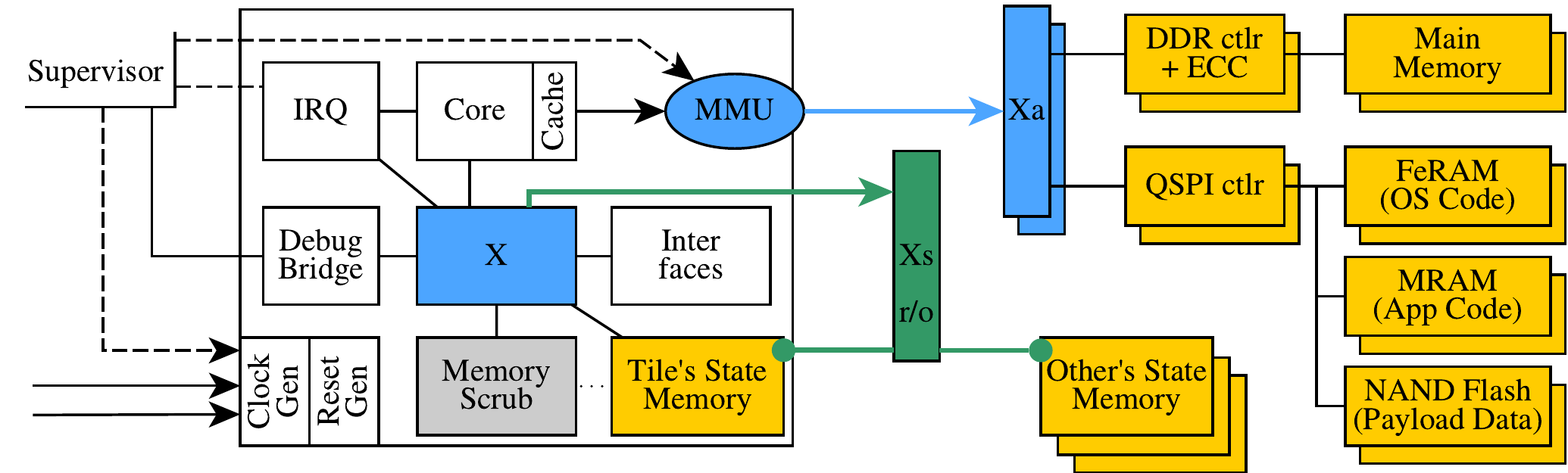}
		\caption{The logic-side architecture of a tile.
			Access to local IP bypasses the cache, while access to global memory passes is cached for performance reasons.}
		\label{fig:tile}
		\vspace{-10pt}
	\end{figure*}
	
	\subsection{Supervision \& Reconfiguration}
	Stage 1 can be implemented on a single chip, but we utilize an off-chip supervisor to facilitate FPGA reconfiguration and transient fault scrubbing in the running configuration.
	The outlined multi-stage FT approach puts only minimal load on the supervisor, and it can thus be again implemented using a traditional radiation hardened or tolerant microcontroller.
	The FeRAM-based TI-MSP430FR family would be a solid somewhat radiation-tolerant but non-FT substitute, which is today widely used aboard a broad variety of CubeSats and low-performance COTS products designed for nanosatellite use.
	The level of performance offered by such microcontrollers is usually sufficient only for educational CubeSats and federated systems.
	However, a supervisor in our architecture only receives the majority voting results from the coarse grain lockstep, controls the FPGA, and facilitates reconfiguration through an ICAP controller in static logic.
	Hence, the low level of performance of an MSP430FR, for example, is sufficient, and allows an ultra-low-cost implementation of our approach for academic CubeSat projects and scientific instrumentation.
	
	We deployed configuration error mitigation through Xilinx SEM in combination with supervisor-side scrubbing to safeguard logic integrity.
	However, SEM and scrubbing only detect faults in specific components of the FPGA fabric (e.g. not in BRAM), leaving significant parts of the design unprotected unless logic-side ECC is used.
	
	These measures alone, thus, do not provide sufficient protection for fine-feature size FPGAs.
	Thus, our software-FT functionality can locate faults in the partition of a specific tile, allowing the supervisor to resolve them using reconfiguration.
	We place tiles in separate configuration partitions to enable partial reconfiguration of individual tiles, without affecting the rest of the system.
	
	As depicted in Fig. \ref{fig:supervision}, the supervisor only reacts to disagreement between tiles, otherwise remaining passive.
	It maintains a fault-counter for each tile and acts as a watchdog.
	When resolving transient faults within a tile, it increments the fault-counter and induces a state update through a low-level debug interface.
	After repeated faults, the supervisor will replace the tile by adjusting the thread-mapping of a spare tile, activating it, and rebooting the faulty tile.
	In case a system developer indicated threshold is exceeded, the disagreeing tile is assumed permanently defunct and not re-used as a spare.
	
	To allow supervisor access to a tile and its address space, each tile is equipped with an AXI debug-bridge (Fig. \ref{fig:tile}).	
	The supervisor can trigger execution of self-test functionality within a tile to detect faults in peripherals.
	It can also trigger an adjustment of a tile's thread allocation as part of Stages 1 and 3, making the MPSoC's computational performance, robustness and energy consumption adjustable at runtime.
	
	Majority voting between tiles can be implemented as distributed majority decision \cite{katta2015ravana}, then requiring no direct intervention of the supervisor during regular operation.
	If this is not desired, or lockstep through interrupt triggered checkpoints is implemented, then the supervisor should also take care of receiving the voting results generated on each tile.
	In that case, the supervisor can access each tile's thread mapping via each tile's debug interface, and if necessary induce a reset or otherwise manipulate a tile without requiring its cooperation.
	
	\begin{figure}[!b]
		\centering
		\includegraphics[width=1.0\linewidth,clip, trim = 0pt 0pt 0pt 0pt]{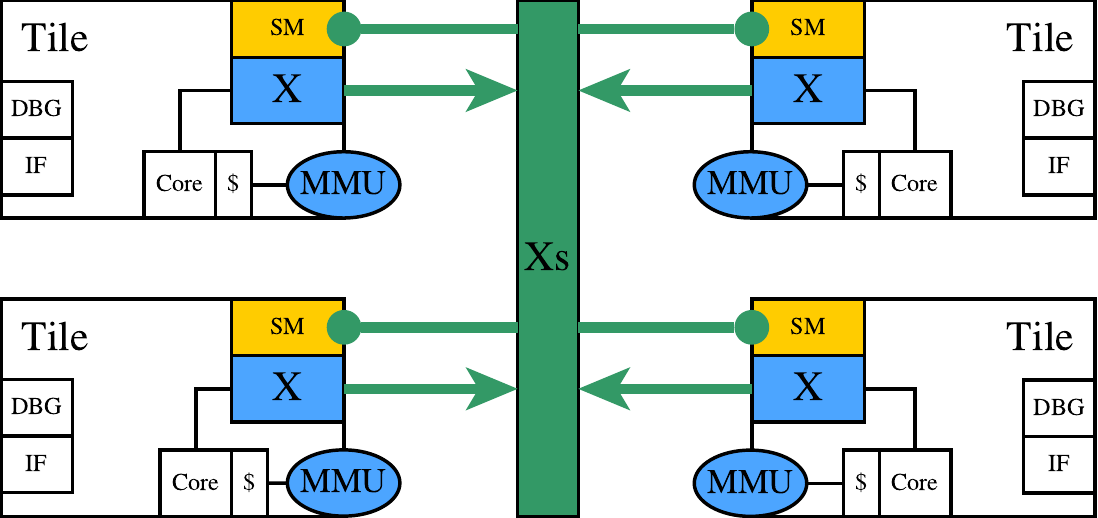}
		\caption{A tile's state memory is accessible to all other tiles in the system.
			It provides a write protected, high-speed on-chip possibility to expose state-relevant data to the MPSoC as a while.}
		\label{fig:statememory}
	\end{figure}

	\subsection{Tile Architecture}
	Our MPSoC design implements multiple isolated SoC-compartments accessing shared main memory and OS code.
	Even though the purpose and function of these compartments is different, the topology resembles a tiled architecture instead of a conventional MPSoC design, in which cores share infrastructure and peripherals.
	This topology allows to maximize Stage 1's fault-coverage capacity and allows task mapping for general-purpose software.
	Each such tile contains a processor core, local interconnect, and peripheral IP-cores and interfaces as depicted in Fig. \ref{fig:tile}, resides in its own clock domain, and can be reset independently.
	Allocating a clock domain to each tile improves timing, and reduces logic-overlap and interdependence between tiles.
	Furthermore, we can then also utilize partial reconfiguration and frequency scaling for each tile, as well as clock gating.
	
	A tile executes a set of thread replicas, and its loss can be compensated by the rest of the system.
	To assure a failed tile can not cause performance degradation in the rest of the system (e.g., by continuously accessing DDR or program memory), it can be disconnected off from the global interconnect by the supervisor.
	Non-masked faults (due to radiation, aging, and wear) disrupt the data or control flow of the software running on a tile.
	Stage 1 builds upon this capability at the thread-level, as state differences can be detected by other tiles and often even by the malfunctioning tile itself \cite{fuchs2017atATS}.
	
	All tiles are equipped with an identical set of peripheral interfaces, with controllers being mapped to identical locations and address ranges.
	The tile address space layout is uniform across the system and tiles are indistinguishable for software.
	Hence, application code and data structures are portable between tiles, simplifying thread migration drastically.
	This allows us to reduce the computational cost and complexity of software-lockstepping.

	Thread allocation and information relevant to the coarse-grain lockstep is stored in a dedicated dual-ported on-chip BRAM on each tile.
	This component is denoted as state memory -- SM -- in the figures.
	One port is accessible to the tile's processor core, while the other is read-only accessible to the system.
	This allowing low-latency information exchange between tiles without requiring inter-tile cache-coherence or main memory access.
	The state memory architecture is depicted in Figure \ref{fig:statememory}.
	The supervisor can access and modify each tile's state memory through its debug interface on each tile.
	
	\subsection{Interconnect Topology \& Shared Memory}
	\label{sec:interconnect}
	Figure \ref{fig:mpsoc} depicts the MPSoC's high-level topology.
	Our MPSoC design utilizes an AXI interconnect in crossbar mode to allow tiles access to shared main and non-volatile memory controllers, though we are currently reworking our MPSoC to instead use a NoC \cite{beechu2017hardware}.
	
	Main memory is shared between tiles, as SD- and DDR memory controllers are too large and require too much I/O to instantiate for each tile.
	Each tile has full access to a segment of main memory, which is mapped to the same address range on all tiles (the MMU component in the figures).
	All tiles can access main memory read-only to simplify state synchronization and IPC.
	The supervisor can access each set of main memory controllers directly.
	
	For nanosatellite missions to LEO, often only SECDED ECC support is required and readily available in library IP already \cite{baker2016design}, while basic error scrubbing can be facilitated in software.
	For critical, deep-space, and long-term missions, block coding should be used instead to compensate for the increased impact of SEEs and higher likelihood of MBUs in high-density SDRAM.
	Reed-Solomon ECC as well as error scrubbers are available commercially, or can be assembled from open-source IP.
	The main memory scrubbers are controlled by the supervisor to avoid potential interference by malfunctioning tiles.
	ARM Cortex-A53 as well as Microblaze caches and several local memories and buffers offer ECC support as basic functionality \cite{baker2016design}.
	
	To safeguard main memory, FeRAM \cite{zhang2015single}, MRAM \cite{Tsiligiannis2013}, and mass memory from SEFIs, as well as permanent failure, these memories, their controllers, and their AXI interconnects are implemented redundantly to enable fail-over.
	This also enables further protective measures \cite{fuchs2015dasia}, and allows load distribution for timing critical main memory through segment interleaving.
	Thereby the available DDR memory bandwidth is increased and the overall latency for memory access can be reduced.
	This also enables us to recover an instance of a memory controller on short notice without requiring the full system to be halted\footnote{Note that depending on the used OS, a reboot of a tile may be required. Linux supports modifications to the memory layout and relocation, while simpler OS, such as RTEMS, do not currently know such functionality.}.

	Tiles compete for DDR memory access.
	As our architecture is implemented on FPGA, the clock frequency of each tile's processor core is lower as on ASIC implemented MPSoCs.
	In consequence, the global interconnect as well as DDR memory controllers offer abundant throughput at drastically higher clock frequencies.
	Each processor core caches access to shared memory, drastically reducing the strain on the memory subsystem\footnote{Access to a tile's state memory still bypasses the cache, but this is implemented directly in high-speed, low-latency on-chip BRAM}.
	Hence, while in principle competing for memory bandwidth, even an 8-tile system can not saturate the two available DDR4 channels in our current MPSoC design.	
	Ideally however, our architecture should be implemented using a NoC instead of a global AXI-interconnect crossbar, which would offer drastically better scalability, more effective caching and buffering, and also a degree of FT.

	\section{Subsystem Connectivity and Peripheral I/O}
	A fault resolved in Stage 1 may cause incorrect data to be emitted through I/O interfaces.
	This is an inherent limitation of coarse-grain lockstep concepts, and can only be slightly alleviated through additional application-intrusive work-around as described, for example, in \cite{dobel2014operating}.
	Instead, this limitation is better solved at the logic level through interface-level voting, which is possible with minimal extra logic.
	For most CubeSats, most nanosatellites, and less critical microsatellite missions, however, this is usually foregone.
	
	Larger spacecraft already utilize interface replication or even voting to assure full hardware TMR, usually requiring considerable effort in hardware or logic to facilitate this replication.
	Our MPSoC architecture inherently provides interface replications by design, requiring no extra measures to be taken, as the individual tile-interfaces can be directly used for TMRed architecture.
	Further safeguards are necessary for very small CubeSats where interface replication is undesirable, for example, due to PCB-space constraints.
	
	\subsection{Electrical- and Logic-level Interface Voting}
	For simple embedded interfaces like I2C and SPI connected to ``dumb'' sensors or actuators with no user configurable firmware, a simple majority decision per I/O line is possible.
	While hardware voting is challenging for large arrays of voters running synchronized at very high frequencies, the CubeSat-relevant interfaces are electrically simple, have a very low pin count, and run at relatively low clock frequencies.
	Hence, voting for these interfaces can efficiently be implemented on-chip through simple voters assuming tiles signals interface activity.
	
	Our coarse grain lockstep mechanics allow software to be executed with slight timing variations.
	These may be caused by clock-domain interactions, competition of tiles for global interconnect DDR4 and QSPI access, as well as differences in tile partition routing and or I/O pin placement.
	In general, these variations will be limited to few clock cycle duration.
	I/O on these interfaces must be buffered, which can be done within the FPGA as discussed further also by Li et al. in \cite{li2010synchronization}.
	For simplicity, tiles should also indicate that an interface is active, and we can double-use the chip-select pins present in almost all I2C and SPI implementations.
	The voter can use activity on these pins as indication that the interfaces is active, and delay voting for a given amount of clock cycles using a set of FIFO buffers.
	The depth of these FIFOs thereby determines the maximum delay compensated by the voter \cite{standeven1989hardware}.
	
	Note that larger MPSoC variants with 6 or more tiles can host multiple independent lockstep sets as described in \cite{fuchs2018atAHS}.
	In this case, simple buffered voting is insufficient, as tiles could then also run mixed lockstep groups where threads may be scheduled with much larger time differentials.
	This differential will always be shorter than the duration of a lockstep cycle or the frame time, but in LEO these may extend to up to several seconds.
	It would be uneconomical and, depending on the application, even technically infeasible to buffer I/O for long duration.
	However, we consider the design-combination of a low-end CubeSats that can not afford subsystem TMR, packet-based communication, with a high-performance 6-core MPSoC not very attractive and therefore a corner case.
	If this combination was still deemed necessary, a straight forward solution would be to maintain multiple isolated thread-assignment groups.

	\subsection{Inter-Subsystem and Controller Networks}
	Many SPI and I2C implementations support multi-master shared bus operation, and it is possible to even create large and complex CAN-bus networks \cite{tai1999cots}.
	CubeSats often use these interface standards for low-speed inter-subsystem communication in simple CubeSat designs \cite{kimm2014controller, bouwmeester2017survey}.
	While packet based interfaces offer far better scalability, reliability, and fault-mitigation properties for this purpose \cite{wilson2016mucsp}, in reality these concepts will remain in use aboard CubeSats for the foreseeable future.
	However, in contrast to interfacing with ``dumb'' endpoints ICs, these networks\footnote{In CubeSat jargon often referred to as ``buses''.} usually consist of microcontrollers running satellite developer provided software.
	In this case, a better solution to de-replicating and obtain consensus within the system of our MPSoC's tiles is to make the subsystems aware of the replication.
	
	A subsystem controller then can await receiving a second replica of a command sequence from a different master.
	Of course this does not solve the issue of a single tile/master jamming or saturating the bus due to malfunction.
	However, most CubeSats using these interfaces as subsystem-bus currently usually also do not take actual meaningful countermeasures in this regard.
	This is technically possible, but requires entirely different network topologies \cite{tai1999cots, wilson2016mucsp} than the simplistic single-level bus concepts used aboard CubeSats today \cite{bouwmeester2017survey}.
	
	\subsection{Routed and Switched Topologies}
	For packet-based interfaces such as Spacewire, AFDX, CAN, or Ethernet, no hardware- or logic-side solution is necessary.
	There, packet duplication and integrity checking can be managed efficiently at the data link, network and transport layers (OSI layers 2 - 4).
	At the physical layer, Ethernet and thereof derived technologies such as AFDX \cite{AFDX} and TTEthernet \cite{gavrilut2017fault} perform shared medium through collision detection and micro-segmentation with frame switching.
	Then, packet routing (L3) and de-duplication in software at the higher OSI layers can be deployed, e.g. in software.
	Today, this is common practice in relevant industrial applications such as AFDX and TTEthernet used in related fields such as atmospheric aerospace or safety critical automotive applications.

	The FPGAs considered in our research provide an abundance of high-speed GTH/GTY transceivers \cite{brebner2011reconfigurable}.
	These are intended to support high-performance serial interfaces such as PCIe, or USB3 host interfaces, which may become attractive for CubeSat use in the future and have built in error correction support.
	Even the smallest XCKU3P part fields 16 such interfaces, and the location of these interfaces is in very attractive locations for using 2-3 of them isolated within each of our MPSoC's tiles \cite{anderson2016felix}.
	In practice, this would allow for a very scalable, high-performance CubeSat inter-subsystem communication architecture \cite{dreschmann2015framework} at little cost assuming a the satellite's high-level design takes this into account.
	
	\section{Applications}
	\label{sec:applications}
	
	The MPSoC architecture described in this contribution was developed for miniaturized satellite use, as an ideal platform for the software-FT approach described in \cite{fuchs2017atATS}.
	It was implemented on a Xilinx XCKU5P FPGA with modest resource utilization (28\% LUTs, 33\% BRAMs, 16\% FFs, 5\% DSPs) and 1.92W total power consumption with four Microblaze-equipped tiles.
	In this design, tiles were equipped each with one peripheral I2C master controller, one SPI master, as well as a dual-channel GPIO controller, which is rather typical for CubeSat applications, while CAN or Spacewire are today not widely used aboard CubeSats.
	However, in \cite{fuchs2017atATS} we also showed that a tile's logic footprint is relatively small in comparison to a large processor core, caches, or globally shared resources such as the global AXI interconnects and the DDR memory controllers.
	Hence, the peripherals allocated to each tile is mostly relevant in terms of the I/O resources required, not regarding the logic footprint.
	We have also developed a variant of our proof-of-concept for the smallest Kintex Ultrascale+ part \texttt{XCKU3P-SFVB784-1LV-I}, and could there reduce the energy consumption of the system to 1.78W.
	
	This architecture is not specifically dependent on utilizing ARM processor cores, but can be implemented with any FPGA-implementable core.
	Our choice of the ARM platform was taken in part to allow thread migration between soft- and hard-cores (e.g., on Zynq Ultrascale+), maximum comparability to COTS mobile-market and embedded MPSoCs with secondary use aboard a major share of CubeSats.
	Especially for low-budget CubeSat users in research or university projects, standard vendor library cores such as Xilinx Microblaze may be an excellent alternative to our Cortex-A choice.
	These cores offer erasure coding and other basic fault-tolerance features out of the box already, and performed rather well in radiation tests \cite{baker2016design}.
	They are readily available and often even free of charge, especially to academics and non-commercial scientific research users.
	The relaxed cost, energy, and size constraints aboard microsatellites and larger spacecraft allow an implementation of our MPSoC spanning multiple FPGAs.
	A multi-FPGA MPSoC variant offers better scalability due to easier routing, can tolerate chip-level defects, and SEFIs to the globally shared memory controllers, these can be distributed to different FPGAs.
	Thread replicas can then be distributed across FPGAs, allowing non-stop operation even during full reconfiguration.
	
	This approach and architecture could very well be implemented on ASIC without reconfiguration and Stage 2, and we see this as a ``big-space'' variant of our approach.
	An ASIC implementation offers lower energy consumption, and allows higher clock rates due to reduced timing and shorter paths.
	If manufactured in an inherently radiation hard technology such as FD-SoI \cite{kochiyama2011radiation}, it would be less susceptible to transients and more robust to permanent faults.
	Due to the drastically increased development cost and required manpower, the resulting OBC would not be viable for most miniaturized satellite applications (not anymore ``on a budget'').

	\section{Outlook \& Future Work}
	Having developed a proof-of-concept implementation of our architecture, it must now be subjected to radiation testing to validate it for on-orbit use.
	Before this was possible, each individual component of our architecture first had to be validated separately.
	This has been achieved or proven by fellow researchers for all individual components comprising our architecture except for our software-FT mechanics.
	
	To validate our software-FT mechanics, we conducted a fault-injection campaign to deliver the high level of test-coverage required to assure the effectiveness of our concept implemented in RTEMS.
	We presented early fault injection results of this campaign in \cite{fuchs2018atATS}, demonstrating that the approach is indeed effective and efficient.
	A more detailed test report of this campaign is forthcoming and we hope to publish it in 2019.
	
	We are currently porting our proof-of-concept MPSoC design to the XRTC KU060 Gen4 backplane family, which is under development by the Xilinx Radiation Test Consortium (XRTC).
	We have achieved an implementation on the Kintex Ultrascale and Ultrascale+ as well as Virtex Ultrascale+ FPGA families, and the work port our existing MPSoC implementation to the \texttt{XCKU060-FFVA1517-1-I} is mostly complete.
	However, at the time of writing the pin-assignments and daughtercard connector mappings of the KU060-based backplane are not yet finalized, and we will have to adjust our design to accommodate those changes over the course of 2019.
	This design, then would then also be directly portable to the space equivalent part \texttt{XQRKU060-CNA1509}.
	CubeSat users, however, would with near certainty still prefer to use the industrial-grade XCKU060 or XCKU3P due to their lower cost and their reduced power consumption.
	
	\section{Conclusions}
	The 3-stage FT approach combined with its MPSoC host system presented in this paper is the first practical, non-proprietary, affordable architecture suitable for FT general-purpose computing aboard nanosatellites.
	It utilizes FT measures across the embedded stack, and combines topological with software functionality, utilizing only extensively validated standard parts.
	Thereby, we enable the use of nanosatellites in critical space missions, while the architecture allows trading processing capacity for reduced energy consumption or fault-coverage.
	
	An OBC relying upon this architecture can be facilitated with the minimal manpower and financial resources.
	The MPSoC can be implemented using only COTS hardware and extensively validated, and widely available library IP, requiring no proprietary logic or costly, custom space-grade processor cores.
	It offers a high level of resource isolation for each processor, utilizing architectural features originally conceived for ManyCore systems to achieve FT.
	
	Each tile functions as a stand-alone processing compartment with dedicated I/O, existing in its own clock domain and reconfiguration partition, thereby minimizing shared resources and reducing routing complexity.
	Compartments were purposefully designed to best support thread-level coarse-grain lockstep of weakly coupled cores, while allowing partial reconfiguration without stalling the rest of the system.
	The architecture was implemented successfully, and tested on current generation Xilinx Zynq/Kintex and Virtex FPGAs with 4, 6 and 8 tiles, and validated through fault-injection into RTEMS.

	\section*{Acknowledgment}
	We would like to thank Gianluca Furano, Giorgio Magistrati, Antonios Tavoularis and Kostas Marinis at ESTEC/TEC-EDD and Melanie Berg at the NASA Goddard Space Flight Center for their support and invaluable feedback.
	We thank ARM Ltd. for making available the relevant processor and infrastructure IP. 
	We would also like to thank the members of the Xilinx Radiation Test Consortium for their encouragement, support, and discussions.

	\bibliographystyle{IEEEtran}
	\bibliography{radecs}
	
\end{document}